\documentclass[twocolumn,printnumbers,amsmath,amssymb,showpacs,prl]{revtex4}
\usepackage{graphicx}
\usepackage{colordvi}

\begin{document}
\title{From Crystals to Disordered Crystals: A Hidden Order-Disorder Transition}

\date{\today}

\author{Hua Tong$^1$}
\author{Peng Tan$^2$}
\author{Ning Xu$^{1,*}$}

\affiliation{$^1$CAS Key Laboratory of Soft Matter Chemistry, Hefei National
  Laboratory for Physical Sciences at the Microscale, and Department of Physics,
  University of Science and Technology of China, Hefei 230026, People's Republic
  of China\\
$^2$State Key Laboratory of Surface Physics and Department of Physics, Fudan
University, Shanghai 200433, People's Republic
of China.
}

\begin{abstract}
We find an order-disorder transition from crystals to disordered crystals for
static packings of frictionless spheres. While the geometric indicators are mostly
blind to the transition, disordered crystals already exhibit properties apart from
crystals. The transition approaches the close packing of hard spheres, giving rise
to the singularity of the close packing point. We evidence that both the transition
and properties of disordered crystals are jointly determined by the structural orders
and density. Near the transition, the elastic moduli and coordination number of
disordered crystals show particular pressure dependence distinct from both crystals
and jammed solids.
\end{abstract}

\pacs{61.43.-j,63.50.Lm}

\maketitle

Order and disorder constitute two fundamental themes in condensed matter physics and
materials science. Perfect crystals, the epitome of order, provide an important starting
point for understanding properties of solids \cite{Crystal}.
In contrast, materials such as glasses and granular assemblies are highly disordered, which
exhibit a set of universal properties distinct from their crystalline
counterparts \cite{Amorphous,Binder1,Berthier1,Berthier2,Granular,Granular2,Torquato2}.
Expectedly, a crystal can evolve away from the perfect crystalline order and
develop into an amorphous state when disorder is introduced \cite{Torquato1,Torquato2}.
While intensive efforts have been invested into properties of amorphous solids, the
characterization of the regime between crystals and amorphous solids,
especially the crossover between the physics of crystals and that of disordered solids,
has not been carefully tackled, leaving the boundary between the two extremes of order
and disorder vague.

Intuitively, one may distinguish disordered solids from crystals based on the structural
orders and simply classify solids with high structural orders to crystals. This has been
proven infeasible because some solids with extremely high structural orders could
exhibit features of disordered solids \cite{Good,Kurchan}; whereas those with considerably
low structural orders may respond like crystals \cite{Good}. Moreover, a recent
experiment showed that a glass could exhibit low-temperature thermodynamic properties
like polycrystals when being compressed to high pressures \cite{Density}. This suggests
that in addition to the structural orders there exist other possible controlling
parameters of the manifestation of disordered solids, while the density is a candidate.
However, how the density works together with the structural orders to determine properties
of disordered solids is still an open question.

Bearing in mind these puzzles, we numerically investigate the evolution
from perfect crystals to disordered solids, by successively tuning the particle-size polydispersity.
In the similar framework, previous simulations have characterized the structural amorphisation
at a sufficiently large polydispersity \cite{Barrat1,Barrat2}. Here
we focus on another unknown order-disorder transition at a rather small polydispersity
from crystals to {\em disordered crystals}, namely solids with extremely high crystalline
order in structure but mechanical and vibrational properties resembling disordered solids.
While the structural orders \cite{Torquato1, Torquato2, Bond1,Bond2}
are insensitive to this transition, multiple quantities undergo apparent changes.
The critical polydispersity of the transition $\eta_c$ is scaled linearly
with the packing fraction distance from the close packing of hard spheres,
$\phi-\phi_{\rm cp}$. Therefore, the close packing behaves like a singular point where
infinitesimally small polydispersity turns the crystal into a disordered crystal
\cite{Kurchan}. The significance of this transition is further manifested by some unknown
physics, which unveil important aspects of solids and answers
the questions raised above.

We start with a perfect crystal, i.e., triangular lattice in two dimensions (2D) and
face-centered cubic lattice in three dimensions (3D). Periodic boundary conditions are
applied in all directions. All particles have the same mass $m$ and interact via the potential
\begin{equation}
V(r_{ij})=\frac{\epsilon}{\alpha}\left(1-\frac{r_{ij}}{\sigma_{ij}}\right)^\alpha
\Theta\left(1-\frac{r_{ij}}{\sigma_{ij}}\right),
\end{equation}
where $r_{ij}$ is the separation between particles $i$ and $j$, $\sigma_{ij}$ is the sum
of their radii, and $\Theta(x)$ is the Heaviside function. It is
a valid model of experimental systems such as granular and colloidal solids
\cite{Ning1,Chen,Hanifpour}. The disorder is continuously introduced by
tuning the particle-size polydispersity. Particle $i$ is assigned a random number $x_i$
uniformly distributed in $[-0.5, 0.5]$, which leads to the particle diameter
$\sigma_i=(1+x_i\eta)\sigma$ with $\eta$ the polydispersity. We increase $\eta$ from $0$
successively by a small step size $\Delta \eta \in [1.6\times 10^{-6}, 8\times 10^{-4}]$,
with smaller $\Delta \eta$ applied near the close packing. After each
change of $\eta$, we rescale the average particle radius to maintain a fixed packing
fraction. The system is then relaxed to the local potential energy
minimum by the fast inertial relaxation engine minimization method \cite{Fire}.
To obtain the normal modes of vibration, we diagonalize the dynamical matrix using
ARPACK \cite{arpack}. The mass, energy, and length are in units of the particle mass $m$,
characteristic energy scale $\epsilon$, and average particle diameter $\sigma$. We show
here results of 2D packings with harmonic repulsion ($\alpha=2$). If not specified, the
number of particles is $N=1024$. We have verified that our major findings are valid for
packings with Hertzian repulsion ($\alpha=5/2$), in both 2D and 3D, and with other
polydispersity distributions, e.g., Gaussian.

\begin{figure}[t]
        \includegraphics[width=0.46\textwidth]{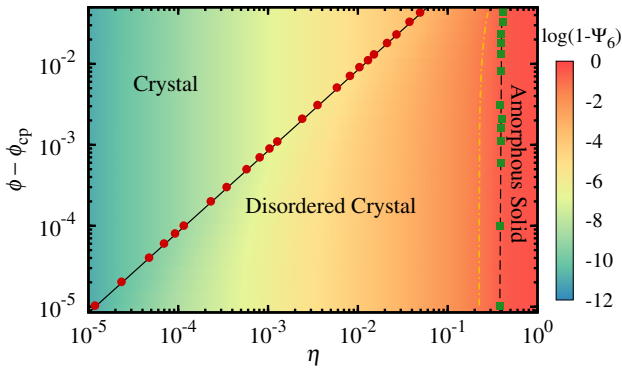}
    \caption{Phase diagram with two order-disorder transitions in
      the parametric space of the polydispersity $\eta$ and
      packing fraction $\phi-\phi_{\rm cp}$. The circles with a linear fit $\eta_c\sim \phi -\phi_{\rm cp}$ (solid
      line) mark the transition from crystals to disordered crystals.  The transition
      labeled by the squares signals the structural amorphisation from
      disordered crystals to amorphous solids (see the Supplemental Material \cite{SupMat} for the definition). The dashed line is to guide the
      eye. The color contour shows ${\rm log}(1-\Psi_6)$ with $\Psi_6$ the bond
      orientational order. The dot-dashed line marks $\Psi_6=0.95$, to the left
      of which all states have a  high crystalline order $\Psi_6>0.95$. }
  \label{fig1}
\end{figure}

Figure~\ref{fig1} shows a unified phase diagram of solids over the entire spectrum of
disorder with both the transitions from crystals to disordered crystals and from
disordered crystals to amorphous solids. Here we denote disordered solids with strong
structural amorphisation as amorphous solids to distinguish them from disordered crystals.

\begin{figure*}
  \begin{center}
    \includegraphics[width=0.95\textwidth]{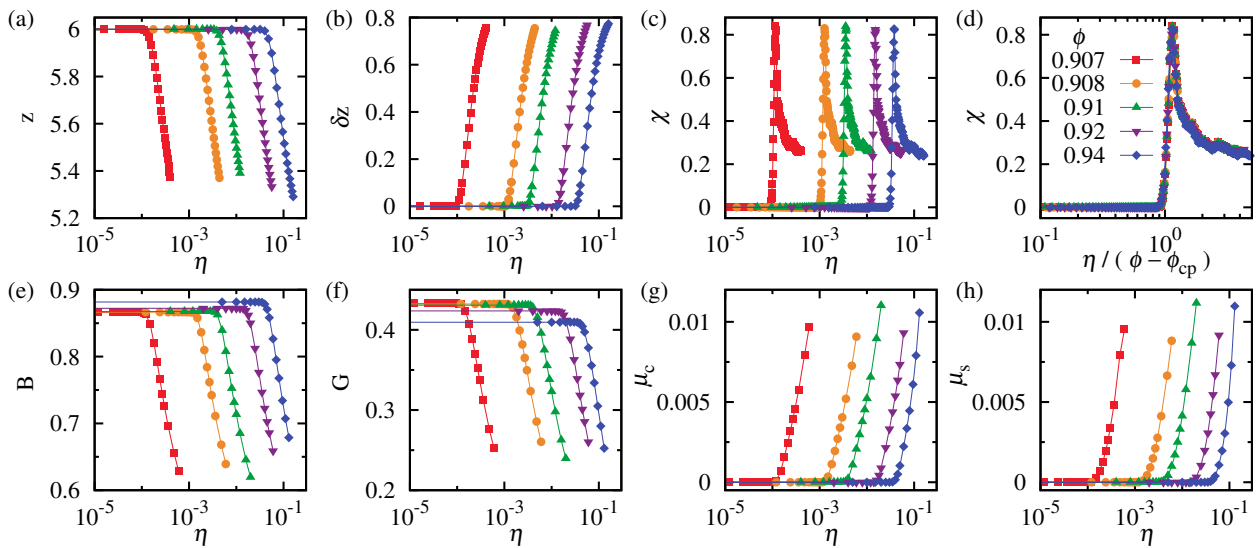}
    \caption{(a)-(c) Polydispersity
        evolutions of the average coordination number $z$,
        fluctuation of the coordination number $\delta z$ as the order parameter, and susceptibility of the order parameter $\chi$. (d) Scaling collapse of all
        curves in (c) when $\chi$ is plotted against $\eta/(\phi-\phi_{\rm
          cp})$. (e)-(h) Polydispersity evolution of the bulk modulus $B$,
        shear modulus $G$, nonaffinity of the compression deformation $\mu_c$,
        and nonaffinity of the shear deformation $\mu_s$. }
      \label{fig2}
    \end{center}
\end{figure*}

When a crystal is driven progressively into a disordered crystal, the
contact network distorts successively and is eventually
destroyed by some local contact breaking. Consequently, the average coordination
number $z$, i.e., the average number of particles with which a given particle
interacts, drops below $6$, as shown in Fig.~\ref{fig2}(a). The contact breaking
happens randomly in space, resulting in the spatially heterogeneous disorder,
which is one of the most important features of disordered solids
\cite{Tong,Binder1,Yoshimoto,Tsamados,Amorphous,Harrowell,Tanaka2}. We thus
propose $\delta z=\sqrt{\frac{1}{N}\sum_{i=1}^{N}(z_i - z)^2}$ as the order
parameter to characterize the strength of disorder, where $z_i$ is the
coordination number of particle $i$. As shown in Fig.~\ref{fig2}(b), $\delta z$
increases quickly from zero at some polydispersity, signaling the transition
from crystals to disordered crystals.

To unambiguously determine the transition point, we calculate the susceptibility $\chi=N[\langle (\delta z)^2\rangle-\langle \delta
z\rangle^2]$, where $\langle . \rangle$ denotes the average over $1000$ distinct
realizations under the same macroscopic conditions. The susceptibility method
is superior in locating phase transition points
\cite{Binder, Melting1,Melting2,Melting3,Yunker}. As shown in Fig.~\ref{fig2}(c), there is a peak in
$\chi(\eta)$, whose location $\eta_c$ is defined here as the transition point from
crystals to disordered crystals. As illustrated by the circles in
Fig.~\ref{fig1}, $\eta_c\sim \phi-\phi_{\rm cp}$, where $\phi_{\rm
cp}=\sqrt{3}\pi/6$ is the packing fraction of close-packed hard disks. The close
packing point is thus singular, because infinitesimally small polydispersity will
trigger the transition.

Interestingly, when $\chi$ is plotted against $\eta/(\phi-\phi_{\rm cp})$ as
in Fig.~\ref{fig2}(d), all curves collapse nicely,
suggesting that $\eta/(\phi-\phi_{\rm cp})$ be a more meaningful parameter in
control of the transition.

\begin{figure}[t]
    \includegraphics[width=0.45\textwidth]{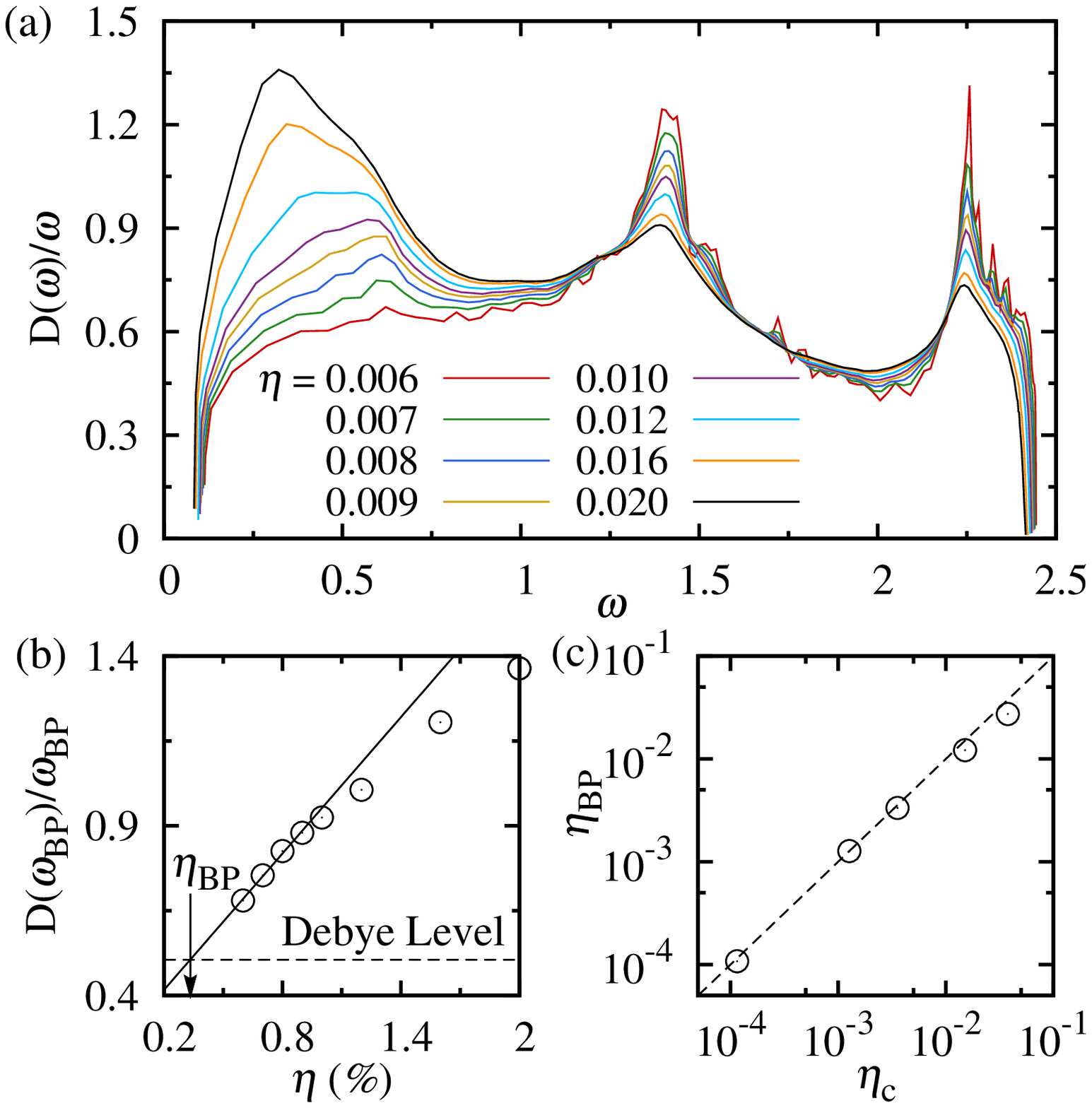}
  \caption{(a) Reduced density of vibrational states
    $D(\omega)/\omega$ at $\phi=0.91$ and
    different polydispersities. From the left to the right, the three peaks are
    respectively the boson peak and two van Hove singularities. (b)
    Polydispersity evolution of the strength of the boson peak
    $D(\omega_{\rm BP})/\omega_{\rm BP}$ from (a). The solid line
    is a linear fit to the low $\eta$ data. It hits the Debye level labeled by
    the horizontal dashed line at $\eta_{\rm BP}$. (c) Correlation between
    $\eta_{\rm BP}$ and $\eta_c$.  The data points are calculated at $\phi=0.907$, $0.908$, $0.91$, $0.92$, and $0.94$ in the
    ascendent order of $\eta_c$. The dashed line shows $\eta_{\rm
      BP}=\eta_c$. }
  \label{fig3}
\end{figure}

Figures~\ref{fig2}(e)-(h) show the $\eta$ evolution of typical properties
characterizing disordered solids, including the elastic
moduli and nonaffinity upon deformation. All quantities change remarkably
across $\eta_c$. In the crystal regime, both the bulk modulus $B$ and shear
modulus $G$ remain mostly constant in $\eta$. Meanwhile, both the compression
and shear deformations are affine with the corresponding nonaffinity
$\mu_c\approx0$ and $\mu_s\approx0$ (see Ref.~\onlinecite{Tong} for the definition of $\mu_c$
and $\mu_s$). When $\eta>\eta_c$, $B$ and $G$ decrease, while $\mu_c$ and
$\mu_s$ increase, all at once. These changes strongly verify the
validity and robustness of the transition.

Figures~\ref{fig2}(e)-(h) also demonstrate that disordered crystals at higher
packing fractions and larger polydispersities (stronger structural disorder) can
have quantitatively similar mechanical properties to those at lower packing
fractions and smaller polydispersities (weaker structural disorder).
In addition to the transition at $\eta_c$, properties of disordered crystals
seem to be also jointly determined by the structural orders and packing fraction
in the form of $\eta/(\phi-\phi_{\rm cp})$. This provides us with some
clues to understand the puzzle why compressed glasses can behave like polycrystals \cite{Density}.
When a glass is compressed, the ratio of the structural disorder
to the density is smaller and approaches the value of polycrystals, which pushes
properties of the glass closer to polycrystals. Therefore, it is the interplay
between the structural orders and the density that determines the performance of
a solid. To claim either of them to be deterministic is partial.

Figure~\ref{fig3} manifest further the importance of disorder to disordered crystals from
vibrational properties. One of the most special vibrational features of disordered solids
is the boson peak, i.e., the peak in the reduced density of vibrational states $D(\omega)/\omega^{d-1}$ with $\omega$ the frequency and $d$ the dimension of space \cite{Amorphous,Binder1,BP1,BP2,BP3,Tanaka5,average}. Figure~\ref{fig3}(a) shows $D(\omega)/\omega$ for disordered crystals at $\phi=0.91$. To
smooth out the planewave-like peaks due to finite size effect, we average
$D(\omega)$ at different system sizes from $N=256$ to $1024$
\cite{average}. With increasing $\eta$, the boson peak (the first
peak at low frequencies) gradually rises and moves to lower frequencies, consistent with the argument that the boson peak is correlated with the
structural disorder \cite{Tanaka5,average,BP2}. There are two other peaks of van Hove
singularities at higher frequencies, whose presence
indicates that the solids are still pretty crystalline in structure and possess
hybridized characters of crystals and disordered solids.

By plotting $D(\omega_{\rm BP})/\omega_{\rm BP}$ against $\eta$ in
Fig.~\ref{fig3}(b) with $\omega_{\rm BP}$ the boson peak frequency, we estimate
below what value of $\eta$ the boson peak disappears. Owing to the very strong
finite size effect close to $\eta_c$, we cannot
obtain smooth enough $D(\omega)$ to resolve the boson peak. By
extrapolating the roughly linear part of the low $\eta$ data, we find that
$D(\omega_{\rm BP})/\omega_{\rm BP}$ hits the Debye level at $\eta_{\rm
  BP}\approx \eta_c$. As shown in Fig.~\ref{fig3}(c), $\eta_{\rm BP}\approx
\eta_c$ over a wide range of packing fractions. The
formation of the boson peak is thus another evidence to distinguish disordered
crystals from crystals.

\begin{figure}[t]
    \includegraphics[width=0.45\textwidth]{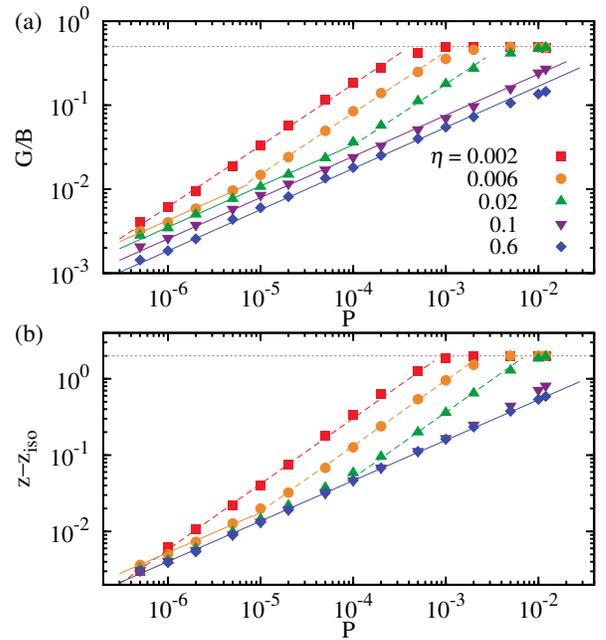}
  \caption{Pressure evolutions of (a) the ratio of the shear modulus
    to the bulk modulus $G/B$ and (b) excess coordination number $z-z_{\rm
      iso}$. The horizontal dotted lines show the crystal behavior. The
    solid (dashed) lines are power law fits to the data: $G/B\sim P^{0.5}$
    ($G/B\sim P^{0.73}$) and $z-z_{\rm iso}\sim P^{0.5}$ ($z-z_{\rm iso}\sim
    P^{0.86}$). }
  \label{fig4}
\end{figure}

Recently, it was proposed that solids spanning the entire spectrum of disorder
could be described by either the physics of jamming or the physics of
crystals \cite{Good}.  For
jammed packings of frictionless spheres, both the excess
coordination number $z-z_{\rm iso}$ and the ratio of the shear modulus to the
bulk modulus $G/B$ are scaled well with the pressure $P$
\cite{jamming0,jamming1,jamming2,jamming3}: $z-z_{\rm
  iso}\sim P^{1/2(\alpha-1)}$ and $G/B\sim P^{1/2(\alpha-1)}$ with $z_{\rm
  iso}=2d$ the isostatic value; while for crystals both
$z$ and $G/B$ are independent of the pressure. With a more accurate control of
structural orders, we check how the two types of physics evolve to each other.

We start from a packing at $\phi=0.92$ and quasistatically decrease the packing fraction
(pressure) at fixed polydispersity. Figure~\ref{fig4} shows the pressure dependence of $G/B$ and
$z-z_{\rm iso}$ at different $\eta$. When $\eta$ is large, the jamming scalings are
recovered over the whole range of pressures studied. When $\eta$ is small, there is a clear transition from the high pressure crystal
scalings to the low pressure scalings disobeying both the physics of crystals and that
of jamming: $G/B\sim P^{0.73}$ and $z-z_{\rm iso}\sim P^{0.86}$. This
transition is right at the transition from crystals to disordered crystals. At
intermediate $\eta$, the three types of scalings are all present, with
the newly reported scalings sitting between those of crystals and
jamming. Therefore, disordered crystals close to the transition at $\eta_c$ comprise a third family of solids complying with the physics other
than crystals and jamming, whose origin and underlying physics are
interesting issues to explore.

The finding and characterization of the transition from crystals to disordered
crystals reveals some unknown features: (i) The close packing is singular
in terms of the transition, implying that it is the only rigid packing of hard spheres satisfying the
physics of crystals;
(ii) the structural orders and density interplay to determine the transition and
properties of disordered solids; (iii) disordered crystals near the
transition exhibit unique pressure scalings apart from crystals and jamming. Here we manifest that our knowledge about solids, even about seemingly crystalline
solids, is still rather incomplete.
Follow-up studies, especially to determine the nature of the transition, are
necessary to have a deeper understanding of
the phenomena reported here.

Our major findings may not be limited to athermal systems with repulsions.
For systems with long-range attractions, we expect
similar results, based on the observation that strong repulsive interactions
govern the behaviors of the packings while attractions act as perturbations
\cite{Ning2}. Since the effective interaction of some widely
studied soft colloidal particles (e.g., poly-N-isopropylacrylamide) can be well fitted to
the repulsions used here \cite{Ning1,Chen}, our observations are also highly
relevant to colloidal experiments, but the temperature effects should be also included. Our preliminary results show that the
transition from crystals to disordered crystals shifts continuously to lower values of the
polydispersity with increasing temperature \cite{tong1}, because the thermal fluctuation is another factor in addition to
the polydisopersity to induce frustrations. Therefore, the athermal case discussed here is not singular from thermal systems and the physics can be generalized to
thermal systems well below the melting temperature, which require further studies and verification of colloidal experiments.

This work is supported by National Natural Science Foundation of China
No. 21325418, National Basic Research Program of China (973 Program)
No. 2012CB821500, and Fundamental Research Funds for the Central Universities
No. 2030020028.

\section{\emph{\textbf{Supplemental Material}}}

\makeatletter
\def\fnum@figure#1{FIG.~S\thefigure$:$~}
\makeatother

\subsection{Section I. Structural amorphisation from disordered crystals to amorphous solids}

  \begin{figure*}
    \begin{center}
      \includegraphics[width=0.9\textwidth]{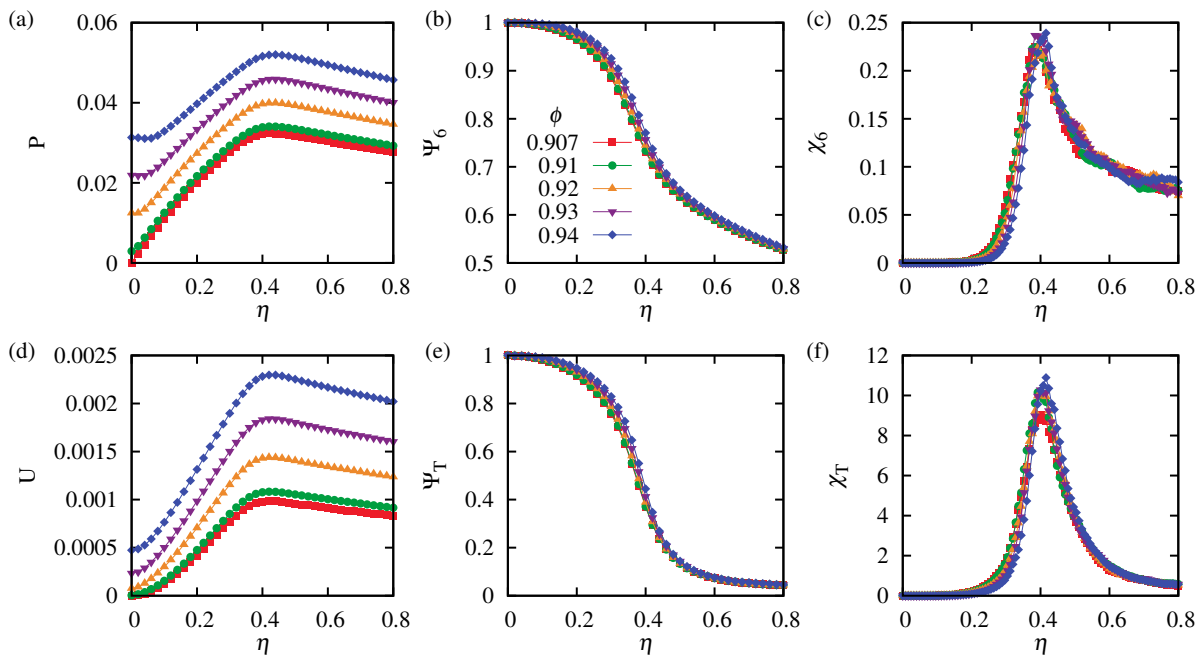}
    \end{center}
    \caption{Polydispersity dependence of ({a}) pressure $P$,  ({b}) bond
      orientational order parameter $\Psi_6$,  ({c}) susceptibility of the
      bond orientational order $\chi_6$, ({d}) potential energy per particle
      $U$, ({e}) translational order parameter $\Psi_T$, and ({f})
      susceptibility of the translational order $\chi_T$. Both $P$ and $U$ show
      a peak when the system undergoes the structural failure. At the same
      point, $\Psi_6$ and $\Psi_T$ show a fast decay and their susceptibilities
      also exhibit a peak. The amorphisation transition signaled by the peak in
      $\chi_{6}$ is shown by the squares in Fig.~1 of the main text.}
    \label{figs1}
  \end{figure*}

As shown in the main text, a crystal undergoes a hidden order-disorder
transition into the disordered crystal phase at a very small particle-size
polydispersity. In this section, we focus on the order-disorder transition at a
sufficiently large polydispersity labeled by the squares in Fig.~1 of the main
text, namely the structural amorphisation \cite{Barrat1,Barrat2}.

Here we calculate two widely used geometric order parameters, the bond
orientational order and translational order \cite{Bond1,Melting2}. The bond
orientational order of particle $j$ is given by
\begin{equation}
\Psi_{6j}=\left|\frac{1}{n_j}\sum_{k=1}^{n_j}e^{6i\theta_{jk}}\right|,
\label{Eq1}
\end{equation}
where $n_j$ is the number of nearest neighbors of particle $j$, and
$\theta_{jk}$ is the angle of the bond between particle $j$ and its neighbor $k$
with respect to the $x$ axis. The global bond orientational order is the average
over all particles: $\Psi_6=\frac{1}{N}\sum_{i=1}^{N}\Psi_{6i}$. The
translational order is defined as
\begin{equation}
\Psi_T=\frac{1}{N}\left|\sum_{j=1}^Ne^{i{\bf G}\cdot {\bf r}_j}\right|,
\label{Eq2}
\end{equation}
where ${\bf G}$ is any first shell reciprocal lattice vector of a hexagonally
close packed system. For a perfect hexagonal lattice, both $\Psi_6$ and $\Psi_T$
are equal to one, while they decrease to small values for highly disordered
solids. In order to unambiguously determine the amorphisation transition point,
we calculate the order parameter susceptibilities
\begin{equation}
\chi_{6,T}=N(\langle \Psi_{6,T}^2\rangle-\langle \Psi_{6,T} \rangle^2),
\label{Eq3}
\end{equation}
where $\langle . \rangle$ denotes the average over $1000$ realizations under the
same macroscopic conditions.

Figure~S\ref{figs1} shows the polydispersity evolution of the pressure $P$,
potential energy per particle $U$, order parameters $\Psi_6$ and $\Psi_T$, and
susceptibilities $\chi_6$ and $\chi_T$ calculated at various packing
fractions. With increasing the polydispersity $\eta$, $P$ and $U$ reach their
peak values at $\eta_a\approx 0.4$. Meanwhile, both $\Psi_6$ and $\Psi_T$ show a
fast decay. The behaviors of $\Psi_6$ and $\Psi_T$ across $\eta_a$ resemble that
of the two-dimensional melting \cite{Melting2}. As demonstrated by
Figs.~S\ref{figs1}(c) and S\ref{figs1}(f), the transition points located by the peaks
in $\chi_6(\eta)$ and $\chi_T(\eta)$ are identical, which signal a one-step
transition from disordered crystals to amorphous solids. The transition of a
(disordered) crystal into a glass, driven by quenched disorder,
has recently been observed in experimental systems of colloidal suspensions \cite{Yunker}.

Note that the critical packing fraction $\eta_a$ of the amorphisation transition
is almost independent of the packing fraction, in contrast to the linear scaling
of the transition from crystals to disordered crystals with respect to the
packing fraction reported in the main text. Moreover, Figs.~S\ref{figs1}(b) and
S\ref{figs1}(e) show that the geometric order parameters as a function of the
polydispersity are also almost independent of the packing fraction, which are
important evidences supporting the generality of Fig.~1 of the main text. If
the polydispersity is replaced with either of the geometric order parameters or
even other structural order parameters, Fig.~1 of the main text is still a
valid phase diagram.

\begin{figure}[t]
  \begin{center}
    \includegraphics[width=0.5\textwidth]{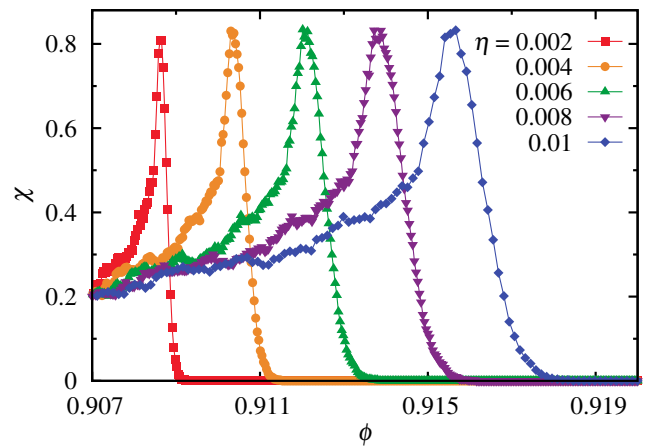}
  \end{center}
  \caption{Susceptibility of the order parameter $\chi$ as a function of the packing
    fraction $\phi$ obtained by decompressing the configurations from
    $\phi=0.92$ at fixed polydispersities $\eta$. The transition points signaled
    by the peaks in $\chi(\phi)$ lie perfectly on the transition line obtained
    on the route of varying $\eta$ at constant $\phi$. }
  \label{figs2}
\end{figure}
\subsection{Section II. Alternate route probing the transition from crystals to disordered crystals}

To demonstrate that the hidden order-disorder transition reported in the main
text is robust and independent of the route, we verify in this section that the
same transition occurs on the route of decompression at fixed particle-size
polydispersity. In Fig.~1 of the main text, this route is perpendicular to that
of varying the polydispersity at fixed packing fraction to obtain the transition
line labeled by the circles. We start from crystal states at high packing
fractions and decompress them at fixed polydispersity by a small decrement of
the packing fraction $\Delta \phi \in [3.625\times 10^{-5}, 2\times 10^{-4}]$,
with smaller $\Delta \phi$ for smaller polydispersity. Potential energy
minimization is performed after each decompression step. The packing fraction
evolution of the order parameter susceptibility $\chi$ defined in the main text
is shown in Fig.~S\ref{figs2}. There is also a peak in $\chi(\phi)$, signaling
the hidden order-disorder transition. We have verified (not shown) that the
transition points obtained here lie perfectly on the transition line (circles)
shown in Fig.~1 of the main text.

\begin{figure}[t]
  \begin{center}
    \includegraphics[width=0.5\textwidth]{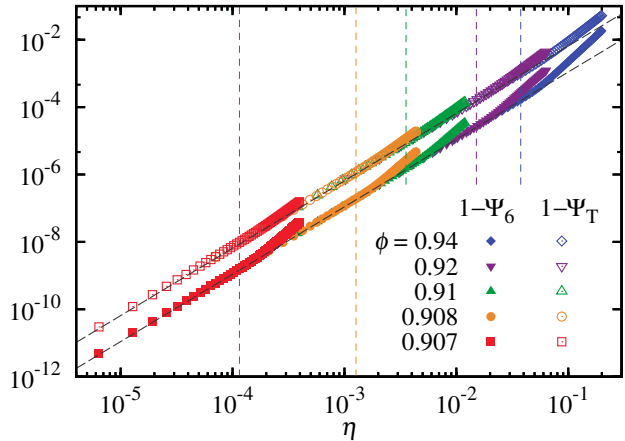}
  \end{center}
  \caption{Over a wide range of packing fractions, the
    deviations from perfect crystalline order of bond orientational order
    parameter $1-\Psi_6$ (filled points) and transitional order parameters
    $1-\Psi_T$ (open points) are plotted as functions of $\eta$. The vertical
    dashed lines label the critical polydispersities of the
    transition. The solid lines have a slope of $2$, indicating the power law
    scalings in the crystal regime. }
  \label{figs3}
\end{figure}

\subsection{Section III. Signs from the geometric order parameters across the hidden
  order-disorder transition }

Seen from Figs.~S\ref{figs1}(b) and S\ref{figs1}(e), there is no observable sign of
the transition from crystals to disordered crystals. However, we still expect to
see some changes of the geometric order parameters across the transition. The
idea is that the mechanical network breaks through the transition, which nontrivially results
in the change of the elastic properties (e.g., the bulk modulus $B$ and shear
modulus $G$ shown in Figs.~2(e)-(h) of the main text) and should also be reflected
in the geometric response. We then plot $1-\Psi_6$ and $1-\Psi_T$ as a function
of the polydispersity and focus on the vicinity of the transition in
Fig.~S\ref{figs3}. In the regime of crystals, both $1-\Psi_6$ and $1-\Psi_T$ are
scaled well with $\eta^2$. Weak deviations from the power-law scalings can be
observed after the crystals transit to disordered crystals.

\begin{figure}[htb]
  \begin{center}
    \includegraphics[width=0.5\textwidth]{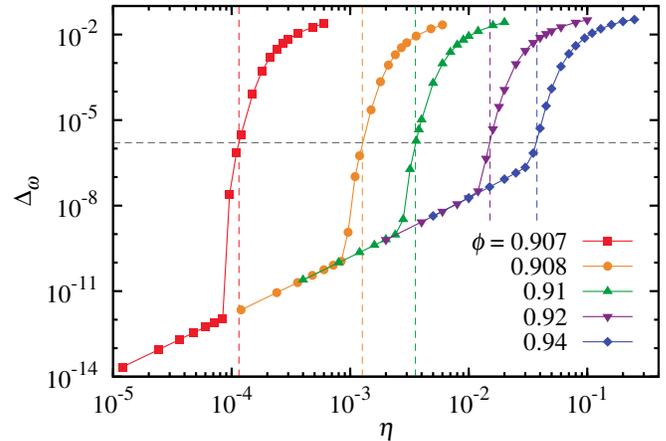}
  \end{center}
  \caption{The departure of the vibrational modes from those of
    the perfect crystal  $\Delta_\omega$ as a function of the polydispersity
    $\eta$. The vertical dashed lines label the critical polydispersities
    $\eta_c$ of the transition
    from crystals to disordered crystals. Interestingly, the critical
    $\Delta_\omega$ at $\eta_c$ are constant in packing fractions, as
    illustrated by the horizontal dashed line.}
  \label{figs4}
\end{figure}
\subsection{Section IV. Another signature of the transition from crystals to disordered
  crystals from the mode analysis}

Normal modes of vibration are the fundamentals to understanding properties of
solids. In a finite-size crystal, the density of vibrational states $D(\omega)$
is composed of a set of $\delta$-functions, with each $\delta$-function
containing degenerate modes. In a weakly disordered crystal, the mode degeneracy
is broken and the $\delta$-functions are broadened with a finite width
positively correlated with the strength of disorder. Therefore, only in the
presence of sufficiently strong disorder and/or for sufficiently large systems,
$D(\omega)$ can be smoothed out. Constrained by the computational power, we are
unable to study large enough systems to reliably resolve the boson peak at small
polydispersities right above the transition from crystals to disordered
crystals. Here we introduce an alternate characterization of the mode evolution
with increasing polydispersity $\eta$, which bypasses the issue of
finite-size effect but shows an interesting and robust feature of the transition
from the perspective of vibrational modes.

We directly trace the evolution of the mode frequencies in reference with those
of a perfect crystal, which is calculated as
\begin{equation}
\Delta_\omega=\left\langle
  \frac{1}{dN-d}\sum_{i=1}^{dN-d}\left(\frac{\omega_i-\omega_i^c}{\omega_i^c}\right)^2
\right\rangle,
\label{Eq4}
\end{equation}
where $d$ is the dimension of space, $\omega_i$ and $\omega_i^c$ are the
eigenfrequencies of the $i$-th mode of the solid with given polydispersity and
of the perfect crystal, and $\langle . \rangle$ denotes the ensemble
average. Due to the periodic boundary conditions, there are $d$ zero-frequency
modes, so the total number of nontrivial modes is $dN-d$. All the frequencies
are sorted in the ascending order. At sufficiently small $\eta$, the $i$-th mode
of a slightly deformed crystal only slightly deviates from the $i$-th mode of
the perfect crystal, so that $\Delta_\omega$ is an exact evaluation of the mode
deviation from a perfect crystal. At larger $\eta$ when mode interchanges take
place, the $i$-th mode of the disordered solid may not be directly evolved from
the $i$-th mode of the perfect crystal any more, so the expected one-to-one
 correspondence breaks. However, we still expect
$\Delta_\omega$ to be a good quantitative calculation of the deviation from
perfect crystals. As shown in Fig.~S\ref{figs4}, $\Delta_\omega$ increases with
$\eta$. In the crystal regime, $\Delta_{\omega}\sim \eta^2$. Near the transition
from crystals to disordered crystals defined by the peak in the order parameter
susceptibility, $\Delta_{\omega}$ grows abruptly. Interestingly,
$\Delta_{\omega}$ at the transition is independent of the packing fraction, as
shown by the intersections between the data curves and vertical lines illustrated
by the horizontal line in Fig.~S\ref{figs4}. This result is suggestive of a universal feature of the
transition from crystals to disordered crystals. In combination with the
observation that the transition at $\eta_c$ may also signal the emergence of the
boson peak, $\Delta_\omega\approx 1.6\times 10^{-6}$ at $\eta_c$ then sets a
critical amount of deviation from a perfect crystal, above which the boson peak
starts to appear. However, the origin of this particular value of
$\Delta_{\omega}$ is unknown at the moment, which deserves further
investigations.

\end{document}